\documentclass[letterpaper, 10 pt, conference]{ieeeconf}
\IEEEoverridecommandlockouts
\usepackage{cite}
\usepackage{amsmath,amssymb,amsfonts}
\usepackage{graphicx}
\usepackage{textcomp}
\usepackage{xcolor}
\usepackage{orcidlink}
\usepackage{algorithm}
\usepackage{booktabs} 
\usepackage{algpseudocode}

\author{Anees Peringal$^{1}$, Mohammed Basheer Mohiuddin$^{1}$, Ahmed Hassan$^{2}$ % <-this % stops a space
\thanks{$^{1}$A. Peringal and M. Mohiuddin are with the Khalifa University Center for Autonomous and Robotic Systems (KUCARS), Khalifa University, Abu Dhabi, UAE. {\tt\small \{100045872, 100059790\} @ku.ac.ae}}
\thanks{$^{2}$ A. Hassan is part of System on a Chip (SoC), Khalifa University, Abu Dhabi, UAE. {\tt\small 100042947@ku.ac.ae}}%
}

\title{ \bf Remaining Useful Life Prediction for Aircraft Engines using LSTM}

\begin{document}

% \author{\IEEEauthorblockN{Anees Peringal}
% % \IEEEauthorblockA{\textit{dept. of Mechanical Engineering} \\
% % \textit{Khalifa university}\\
% % Abu Dhabi, UAE \\
% % 100045872@ku.ac.ae \orcidlink{0000-0002-7851-8542}}
% \and
% \IEEEauthorblockN{Mohammed Basheer Mohiuddin}
% % \IEEEauthorblockA{\textit{dept. name of organization (of Aff.)} \\
% % \textit{name of organization (of Aff.)}\\
% % City, Country \\
% % email address or ORCID}
% \and
% \IEEEauthorblockN{Ahmed Hassan}
% % \IEEEauthorblockA{\textit{dept. name of organization (of Aff.)} \\
% \and
% \and
% \and

\maketitle

\begin{abstract}
This study uses a Long Short-Term Memory (LSTM) network to predict the remaining useful life (RUL) of jet engines from time-series data, crucial for aircraft maintenance and safety. The LSTM model's performance is compared with a Multilayer Perceptron (MLP) on the C-MAPSS dataset from NASA, which contains jet engine run-to-failure events. The LSTM learns from temporal sequences of sensor data, while the MLP learns from static data snapshots. The LSTM model consistently outperforms the MLP in prediction accuracy, demonstrating its superior ability to capture temporal dependencies in jet engine degradation patterns. The software for this project is in \url{https://github.com/AneesPeringal/rul-prediction.git}. 
\end{abstract}

% \begin{IEEEkeywords}
% Remaining Useful Life (RUL), Long Short-Term Memory (LSTM) Networks, Predictive Maintenance, Machine Learning in Maintenance, 
% \end{IEEEkeywords}

\section{Introduction}

Jet engine maintenance and safety in the aerospace industry is a critical area, where Remaining Useful Life (RUL) prediction is paramount for ensuring reliable, safe, and cost-effective operations. Accurately forecasting engine component degradation and potential failure can significantly enhance maintenance strategies and prevent unplanned downtimes. Traditional RUL prediction methods have relied on physical models and statistical approaches, often requiring extensive historical failure data and domain expertise. However, the complex operational environments and intricate failure mechanisms of jet engines pose challenges to these conventional methodologies.

Data-driven approaches, enabled by machine learning and the accumulation of sensor data from engines, have emerged as powerful alternatives. Among machine learning techniques, neural networks have shown promise in RUL prediction due to their ability to model complex, non-linear relationships within data \cite{deng_remaining_2020,ma_predicting_2018,RefWorks:RefID:51-mahamad2010predicting}.

Multilayer Perceptrons (MLPs), while simple and effective for pattern recognition, lack the ability to process sequential data, making them less suitable for time-series predictions where temporal dynamics are crucial \cite{RefWorks:RefID:51-mahamad2010predicting}.

Long Short-Term Memory (LSTM) networks, a specialized form of recurrent neural networks, are designed to address the limitations of MLPs in handling time-series information. LSTMs can learn long-term dependencies and patterns in sequential data, a characteristic inherent to jet engine monitoring data \cite{RefWorks:RefID:70-xia2020data-driven,deng_remaining_2020}.

This study compares LSTM and MLP models for RUL prediction in jet engines. Using the C-MAPSS dataset from NASA comprising sensor readings from a fleet of engines over their operational cycles, we demonstrate the superior ability of LSTM networks in capturing the temporal degradation patterns essential for accurate RUL prediction.

\section{Related Works}

Condition monitoring in machinery maintenance plays a crucial role, primarily achieved through sensors that perform fault prognosis, detection, and diagnosis. These sensors estimate RUL of machinery, a task made challenging due to machinery typically not being operated until failure. Accurate RUL estimations are vital in avoiding premature maintenance and planning component replacements.

RUL estimation utilizes two main approaches: model-based and data-driven. The model-based approach involves collecting data on machinery's state and condition, using it to develop mathematical models that predict future behavior. This approach, often employing state estimators like Kalman filters \cite{RefWorks:RefID:66-cui2020research} or Particle filters \cite{RefWorks:RefID:67-li2021remaining}, necessitates a deep understanding of the machinery's physics and mechanics. Conversely, the data-driven approach leverages historical data from the machinery, applying machine learning algorithms to identify patterns for predictions. This method is less reliant on domain knowledge but requires substantial high-quality data.

Both approaches have limitations. Physical models effectively predict degradation but struggle to generalize RUL estimates. Data-driven approaches are hampered by the scarcity of good data, as machines aren’t typically run to failure. Hybrid models, such as those in \cite{RefWorks:RefID:53-wang2022remaining}, combine both systems’ advantages. These models use physical models for data pre-processing, simulating various degradation models, then fusing the data with LSTM networks.

LSTM-RNNs excel in capturing temporal data patterns, enabling accurate RUL predictions. This methodology, shown to outperform others, still faces challenges, notably the limited generalizability of physical models across different machinery types. An alternative approach by \cite{RefWorks:RefID:55-deutsch2018deep} involved using Deep Belief Networks (DBN) for feature extraction, inputted into Feedforward Neural Networks (FNN). The combination aimed to harness DBN’s feature extraction with FNN’s prediction capabilities, albeit with mixed results compared to other models.

\cite{RefWorks:RefID:51-mahamad2010predicting} explored using FNNs with simple statistical data preprocessing. Data from rotating bearings was processed using the Weibull hazard rate function, with RMS and kurtosis as inputs for the FNN. Although this simple method showed promise, it lacked benchmark data for quality assessment, yet the preprocessing approach proved useful for noisy data.

Particle filters, as an alternative to neural networks, were examined in \cite{RefWorks:RefID:54-naipeng2015improved} and \cite{RefWorks:RefID:52-yaguo2016model-based}. These studies used statistical modeling to filter data and model machinery degradation, with particle filters estimating the machinery state and RUL. While results appeared promising, the authors acknowledged the model’s limitations in handling abrupt, unaccounted changes in degradation.

Degradation models, as inputs to Machine Learning (ML) models, have shown potential, as seen in \cite{RefWorks:RefID:53-wang2022remaining}. Research suggested that the integration of RNNs \cite{RefWorks:RefID:69-mo2021evolutionary,RefWorks:RefID:70-xia2020data-driven, RefWorks:RefID:68-li2019rolling}, CNNs \cite{RefWorks:RefID:73-li2021remaining,RefWorks:RefID:74-mazaev2021bayesian,RefWorks:RefID:71-cao2021temporal,RefWorks:RefID:72-li2019deep}, and support vector machines \cite{RefWorks:RefID:75-manjurul2021data-driven} could adapt to abrupt changes, demonstrating the effectiveness of fusing various ML approaches in machinery health monitoring.

\section{Methodology}

\begin{algorithm}
\caption{Detailed RUL Prediction Using LSTM Neural Network}
\begin{algorithmic}[1]

\State \textbf{Data Preprocessing:}
\State Load training and test data from files.
\State Normalize features using Min-Max scaling.
\State Generate time-series sequences from data for LSTM processing.

\State \textbf{Model Definition:}
\State Define an LSTM neural network with specified layers and units.
\State Initialize weights and biases.
\State Set activation functions for layers 
\State Define a loss function
\State Use Adam optimizer for parameter updates.

\State \textbf{Training Process:}
\State Set the number of epochs for training.
\For{each epoch}
    \State Shuffle the training data to ensure randomness.
    \For{each batch in training data}
        \State Perform forward propagation through the LSTM network.
        \State Compute loss using defined loss function.
        \State Perform backward propagation to compute gradients.
        \State Update LSTM network weights using ADAM optimizer.
    \EndFor
\EndFor

\State \textbf{Evaluation:}
\State Evaluate the trained model on the test dataset.
\State Compute and report performance metrics.

\end{algorithmic}
\end{algorithm}

The C-MAPSS dataset \cite{dataset} consists of training and testing data where the training data has sensor readings recorded till the end of life. The testing data, however, stops recording before the end of life. The objective of this work is to estimate the number of remaining cycles in the engine's life from the end of the sensor recording. 

\begin{figure}
    \centering
    \includegraphics[width = \columnwidth]{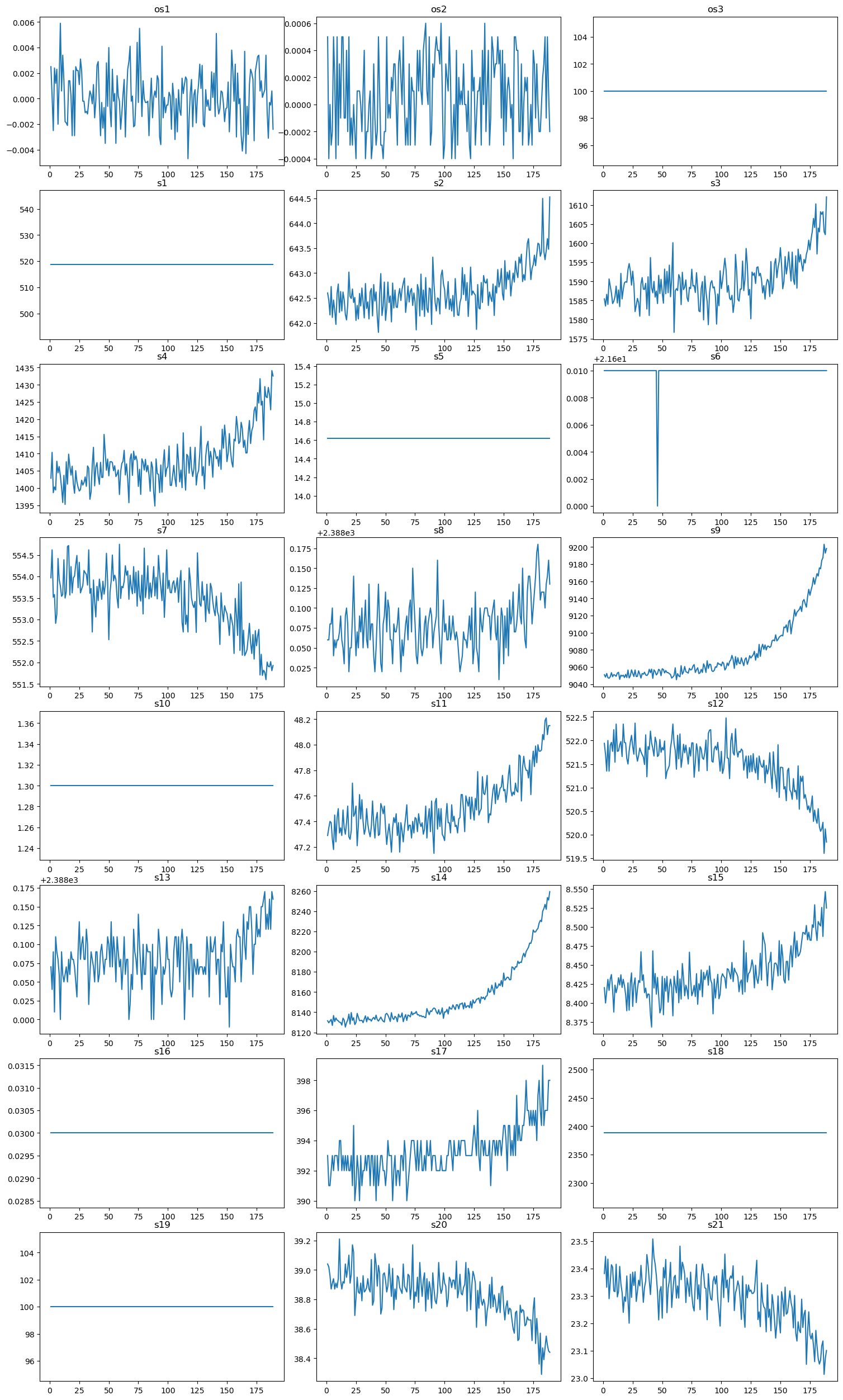}
    \caption{The operating conditions and the sensor readings for the first engine in the dataset are shown here. Some sensors do not provide any useful information for determining RUL.}
    \label{fig:sensors}
\end{figure}

\begin{figure}
    \centering
    \includegraphics[width = \columnwidth]{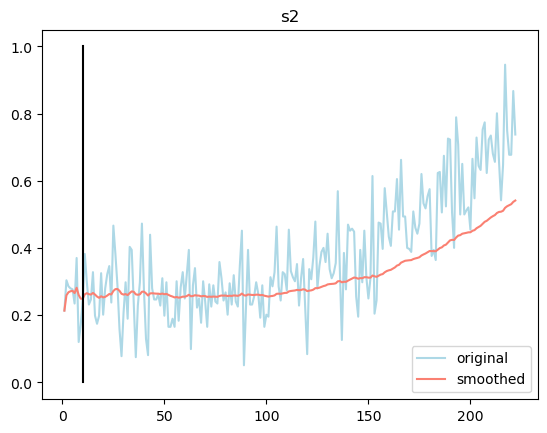}
    \caption{Smoothed sensor readings. We remove the first 10 samples for every sensor because the smoothed signal is not representative of the actual signal in the beginning.}
    \label{fig:smoothData}
\end{figure}

The dataset consists of 100 engines, each having three recorded operating conditions and 26 different sensor readings. The sensors that were used to obtain the reading are not mentioned; therefore, we cannot use domain knowledge of the Turbojet dynamics to predict the RUL. The operating conditions and the sensor readings available in the dataset for a particular engine are shown in Fig \ref{fig:sensors}. Sensors 1, 5, 6, 10, 16, 18 and 19 do not change throughout the operation of the engine. Therefore, they can hardly be used for predicting the remaining useful life. In our subsequent analysis, we do not consider these sensor readings for training the networks using the details as given in Table \ref{table:methodology}. The sensor readings are noisy and we employ exponentially weighted average to reduce the noise in the data, as shown in Fig. \ref{fig:smoothData}.

The valuable sensor data is scaled by min-max scaling \cite{dataset} to keep with $[0,1]$, and the scaling parameters are saved for scaling the data in inference time. Out of the 100 engines in the dataset, we split the data corresponding to 20 random engines as the validation data. This was used during training to ensure that the model does not overfit to the training data.

\begin{table}[h]
\centering
\caption{Training details}
\label{table:methodology}
\begin{tabular}{ccc}
\toprule
\textbf{Aspect}               & \textbf{MLP}                & \textbf{LSTM}               \\ 
\midrule
\textit{Package}              & Pytorch                     & Pytorch                     \\ 
\textit{Epochs}               & 35                          & 35                          \\ 
\textit{Learning Rate}        & 0.001                       & 0.001                       \\ 
\textit{Convergence}          & Fig. \ref{fig:loss_mlp}     & Fig. \ref{fig:loss_curve}   \\ 
\textit{Input Difference}     & Current engine features     & 20-timestep sequences       \\ 
\textit{Data Sampling}        & 64 random points            & 64 sequences                \\ 
\textit{Additional Info}      & No temporal context         & Resets states after batch   \\ 
\bottomrule
\end{tabular}
\end{table}

\section{Results}

\subsection{Training Performance}

\begin{figure}
    \centering
    \includegraphics[width = \columnwidth]{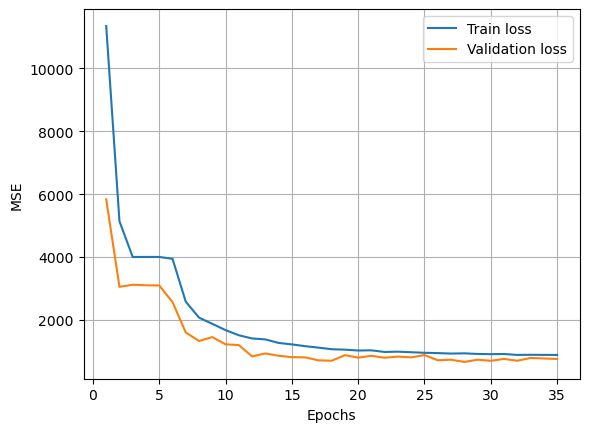}
    \caption{Loss curves for the LSTM model.}
    \label{fig:loss_curve}
\end{figure}

The training progression for the LSTM model, as depicted in Fig. \ref{fig:loss_curve}, was characterized by a steep decline in both training and validation loss, converging to a stable MSE of 796.42. This rapid decline within the initial epochs is indicative of the LSTM's capability to capture the temporal dependencies in the sensor data effectively. The loss curves demonstrate a consistent reduction with negligible overfitting, evidenced by the close tracking of training and validation loss values.

\begin{figure}
    \centering
    \includegraphics[width = \columnwidth]{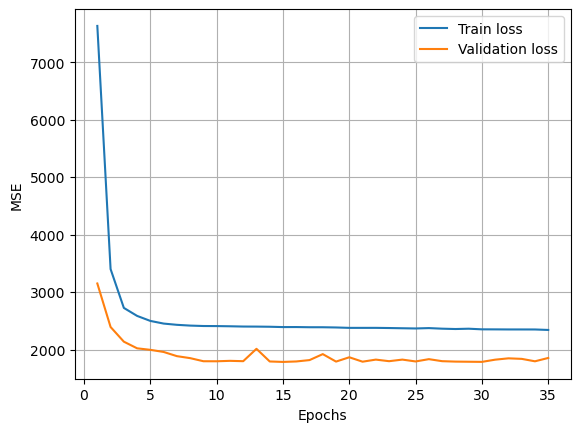}
    \caption{Loss curve for the MLP model}
    \label{fig:loss_mlp}
\end{figure}

Conversely, the MLP model's training, illustrated in Fig. \ref{fig:loss_mlp}, exhibited a less pronounced, yet steady descent in loss values, culminating in a plateau at a higher MSE of 1745. The relative flatness of the curve post the initial epochs points to a potential underfitting scenario, where the MLP fails to model the complexities inherent in the sequential data adequately. The disparity between the training and validation loss indicates a gap in the model's ability to generalize, which is further exacerbated in the validation phase.

\subsection{Prediction Accuracy}

\begin{figure}
    \centering
    \includegraphics[width = 0.45\textwidth]{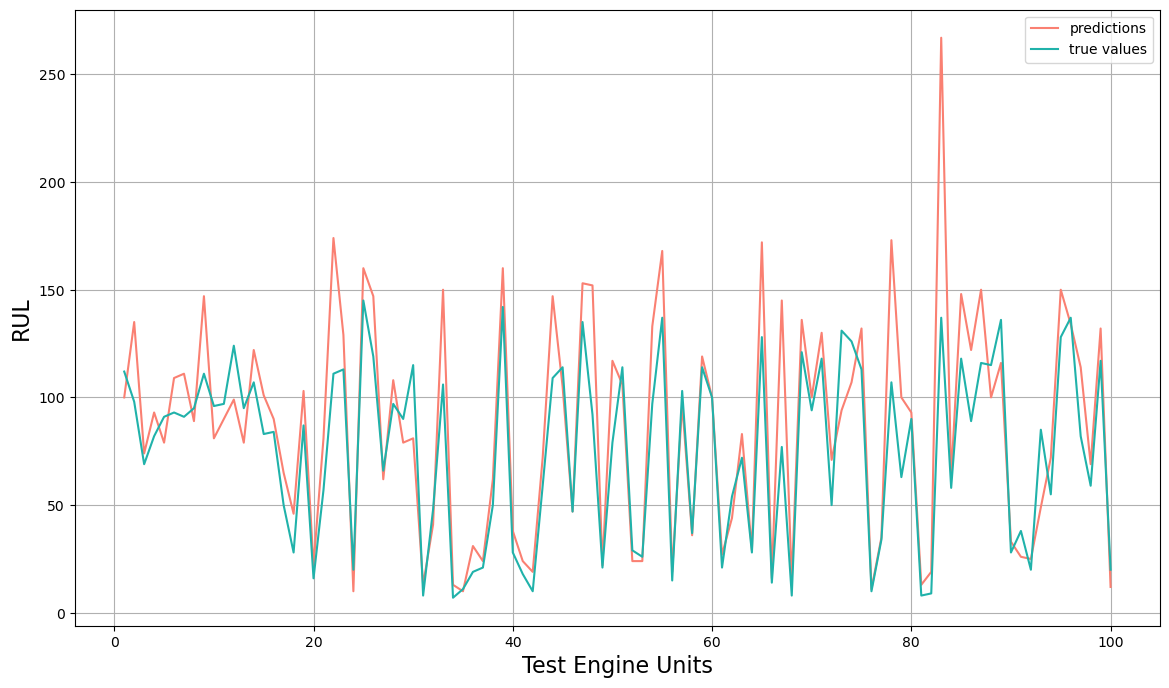}
    \caption{Compares the predicted RUL to the label in the test dataset for the LSTM model. We obtain MSE of 796.42 in the testing set.}
    \label{fig:cmp_RUL}
\end{figure}

The LSTM's capability to predict the RUL of jet engines was further analyzed through its performance on a test set. Fig. \ref{fig:cmp_RUL} compares the predicted RUL values against the true values for different engine units. The LSTM predictions closely follow the true RUL values, demonstrating the model's ability to capture the complex, time-dependent degradation patterns that characterize jet engine life-cycle. While there are instances of deviation, particularly in engines with higher RUL values, the overall trend shows a strong correlation between the predicted and actual RUL. Notably, the LSTM model maintains its predictive performance across various engine units, reflecting its robustness and potential for scalability across different engine types and operational conditions.

\begin{figure}
    \centering
    \includegraphics[width = 0.45\textwidth]{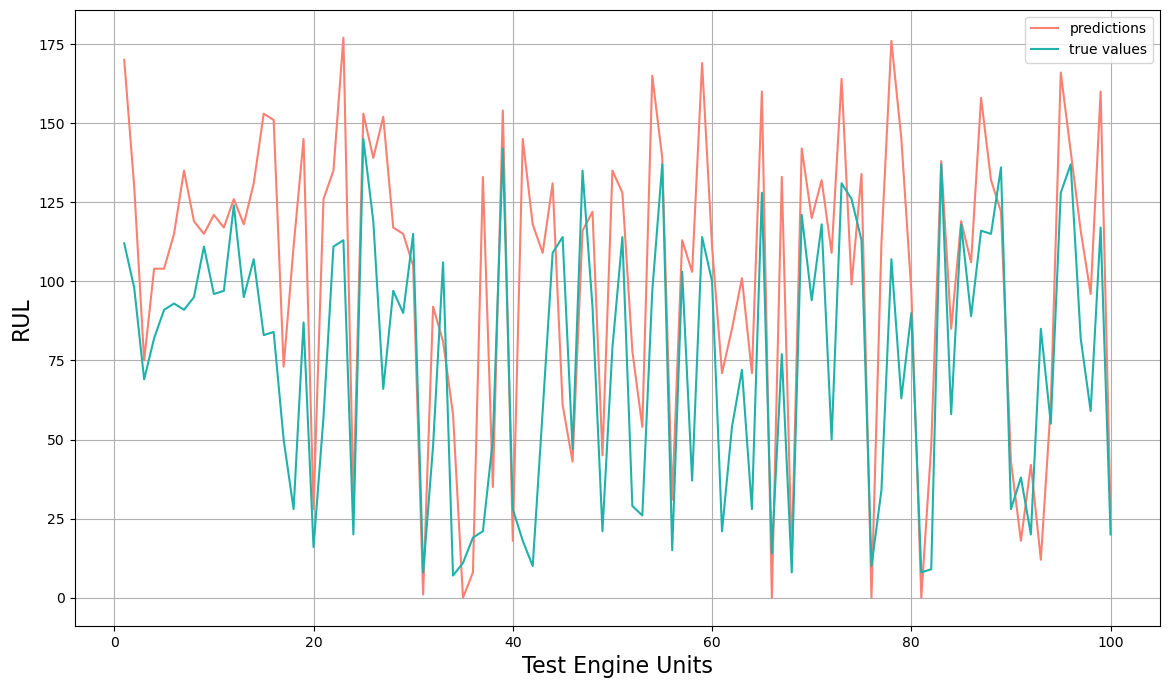}
    \caption{Compares the predicted RUL to the label in the test dataset for the LSTM model. We obtain MSE of 1745.}
    \label{fig:RUL_mlp}
\end{figure}

In contrast, the MLP model's performance, as illustrated in Fig. \ref{fig:RUL_mlp}, displays a more significant deviation from the true RUL values. Notably, the MLP model's predictions exhibit higher variance and a marked difficulty in capturing the rise and fall patterns associated with the engines' degradation curves. This variance is indicative of the MLP's inherent limitations in processing and learning from time-series data, where sequential and temporal dependencies significantly inform the prediction accuracy.

\section{Conclusion}

This study has successfully demonstrated the application of LSTM networks for predicting RUL of jet engines. The LSTM model exhibited a superior ability to learn and predict the complex temporal degradation patterns inherent in the operational data of jet engines when compared to traditional MLP models.

The LSTM's predictive performance, as evidenced by lower MSE and the higher correlation with the true RUL values across various test engine units, holds significant implications for the aerospace industry. By reliably forecasting the RUL, this approach can revolutionize maintenance strategies, shifting from reactive to proactive measures, thereby reducing unexpected downtimes and extending engine lifespans.

\begin{table}[]
    \centering
    \caption{Dataset}
    \begin{tabular}{l l}
      \hline\hline \\
      Data division& FD001 \\\hline
       No. of turbofan engines in training set  & 100\\
       No. of turbofan engines in testing set & 100\\
       No. of types of operating conditions & 3\\
       No. of failure modes & 1\\
       No. of training samples & 17,731\\
       No. of test samples & 100\\\hline\hline
    \end{tabular}

    \label{tab:my_label}
\end{table}

\bibliographystyle{ieeetr}
\bibliography{mybibliography}
% \printbibliography
\end{document}